\documentstyle[epsf,psfig]{l-aa}
\begin{document}
\def\kms{km\thinspace s$^{-1}$}
\def\ergcs{erg cm$^{-2}$ s$^{-1}$}
\thesaurus{11.19.2; 11.11.1; 11.16.1; 11.19.6; 11.09.2}

\title{Effects of interaction on the properties of spiral galaxies. II. Isolated galaxies:
The zero point
\thanks{Based on data obtained at the 1.5m telescope of the Estaci\'on de Observaci\'on de
Calar Alto, Instituto Geogr\'afico Nacional, which is jointly operated by the Instituto
Geogr\'afico Nacional and the Consejo Superior de Investigaciones Cient\'{\i}ficas through the
Instituto de Astrof\'{\i}sica de Andaluc\'{\i}a}}

\author{I.~M\'arquez \inst{1} \thanks{Visiting Astronomer, German-Spanish Astronomical Center, Calar
Alto, operated by the Max-Planck-Institut f\"ur Astronomie jointly with the Spanish National
Commission for Astronomy}
\and
M.~Moles \inst{2,3} $^{**}$ }

\offprints{I.~M\'arquez, isabel@iaa.es }

\institute{ Instituto de Astrof\'{\i}sica de Andaluc\'{\i}a (C.S.I.C.), Apdo. 3004, 18080
Granada, Spain
\and
Instituto de Matem\'aticas y F\'{\i}sica Fundamental (C.S.I.C.), C/ Serrano 123, 28006 Madrid,
Spain
\and
Observatorio Astron\'omico Nacional, Madrid, Spain }

\date{Received,  ; accepted,}

\maketitle

\markboth{M\'arquez \& Moles: Isolated spiral galaxies: II. The zero point}{}

\begin{abstract}
We analyse the properties of a sample of 22 bright isolated spiral galaxies on the basis of Johnson B,V,I
images and optical rotation curves. The fraction of early morphological types in our sample of
isolated galaxies (or in other samples of non-interacting spiral galaxies) appears to be smaller 
than in samples including interacting systems. The overall morphological aspect is regular and 
symmetric, but all the galaxies present non-axisymmetric components in the form of bars or rings. 
We find that the color indices become bluer towards the outer parts and that their central values
are well correlated with the total colors. The properties of the bulges span a larger 
range than those of the disks, that thus are more alike between them. None of the galaxies shows 
a truncated, type II disk profile. It is found that the relation between surface brightness and 
size for the bulges, the Kormendy relation, is tighter when only isolated galaxies are considered. 
We find a similar relation for the disk parameters with an unprecedented low scatter.

A Principal Component Analysis of the measured parameters shows that 2 {\sl eigenvectors}
suffice to explain more than 95 $\%$ of the total variance. These are, as found for other
samples including spiral galaxies in different environmental situations, a scale parameter
given by the mass or, equivalently, the luminosity or the size; and a form parameter given by
the bulge to disk luminosity ratio, B/D, or, equivalently, by the gradient of the solid-body
rotation region of the rotation curve, the G-parameter. We report here a tight correlation
between G and B/D for our sample of isolated spirals that could be used as a new distance
indicator.


\end{abstract}

\section{Introduction}

It is generally admitted that gravitational interaction can modify the properties of galaxies
in rich environments. Its effects are often invoked to explain different observed properties,
from the distribution of morphological types in clusters of galaxies to the peculiarities
sometimes seen in particular galaxies. Those effects, on the other hand, can be very diverse in
nature and have very different time scales to manifest, so it is not straightforward to
ascertain whether such or such peculiarity is actually the fact of the interaction. In other
words, the absence of peculiarities in a given system cannot be given as a sign of isolation,
whereas the presence of unusual features cannot be unambiguously given as a proof of
gravitational interaction (see Moles et al. 1994, for the pair NGC~450/UGC~807).

It is clear that the characterization of the specific effects of the gravitational interaction
needs to be preceded by the exhaustive analysis of the properties of galaxies that could be
considered as isolated. The average values and the ranges they present in size, luminosity,
bulge to disk ratio, etc, do constitute the starting point to which refer similar properties of
spirals in richer environments, from isolated pairs and small groups to clusters. Only such a
comparative analysis could eventually lead to the identification of the specific effects of the
gravitational interaction. This interaction is expected to produce changes in some
morphological aspects, the kinematics, and the stellar content of the involved galaxies. Thus,
it is necessary to start with the study of those same properties for a well defined sample of
isolated galaxies.

Important studies of samples of spiral galaxies do exist, but these, even if sometimes defined
as containing normal or non-peculiar galaxies, include galaxies belonging to interacting
systems. The first important contribution to the study of galactic kinematics was made by Rubin
and collaborators (Rubin et al. 1991, and references therein). A total of about 60 galaxies,
selected to cover a wide range in size, mass, and luminosity were observed. There was no aim to
build up a complete sample, and most of the objects are non isolated. Moreover, the ulterior
photometric analysis (Kent 1988, and references therein) has been done only for some of them,
so complete data are only available for a relatively small number of field spirals. Further
studies have considerably increased the number of objects, but none of them took into account
the information on the environmental status of the galaxies, and were focused either on the
photometric properties (de Jong \& van der Kruit 1994; de Jong 1996a, b, c; see Table 1 from
H\'ereaudeau \& Simien 1996; Peletier \& Balcells 1997; more recently, Baggett et al.  1998 for
one photometric band data) or to spectroscopic and imaging surveys of field galaxies for which
the existing information is long slit spectra together with just one broad band (Mathewson et
al. 1992; Courteau 1996).

It is not a simple question to define what an isolated galaxy is. We only pretend here to
establish operational criteria to identify isolated systems. The perturbations that a galaxy
can suffer depend, apart the properties of the galaxy itself, on the mass, size, distance and
relative velocity of the perturbing agent. Thus, the influence of very far away big galaxies
will be negligible, but small galaxies can produce secular alterations on the dynamics of the
primary system provided they are close enough (Athanassoula 1984; Sundelius et al 1987; Byrd \&
Howard 1992). And they can manifest themselves on very different time scales. Here we define an
isolated galaxy as that for which the possible past perturbations by neighboring galaxies, if
any, have been completely erased by now. Accepting that typical time scales for the decay of
the perturbation effects are not longer than a few times $10^9$ years, a criterium for
isolation can be given. As discussed in M\'arquez and Moles (1996; hereafter paper I), we
consider a galaxy isolated when it has no neighbours in a volume defined by a radius of 0.5 Mpc
in projected distance and a redshift difference of 500 km/s. To be conservative, we also
discarded all those galaxies which appear on the POSS prints with close neighbours for which
there is no redshift information (see below).

The present work, the second of three, is devoted to the study of the properties of a sample of
isolated spiral galaxies. The case of spirals in isolated pairs will be presented in Paper III,
whereas the description of both samples, the details of the observations, data reduction and
methods of analysis were given in Paper I. We have both, CCD multi-color (Johnson B, V and I
bands) photometry and major axis long slit spectra information, for 15
isolated spiral galaxies. For some of those galaxies we also present minor axis long slit
spectra and/or H$\alpha$ CCD photometry. We also present long slit spectra for 4 more galaxies.
The properties we have measured are compared with those of other analysis to find whether they
are different. This contribution is organized as follows: The sample is briefly described in
section 2. In section 3 we comment the morphological aspects. In section 4 we analyse the set
of parameters obtained from the whole data and in section 5, the relationships among them. The
conclusions are presented in section 6.

\section {The sample of isolated galaxies}

In configuring our sample of isolated spiral galaxies, the first step was to select all the
spirals brighter than $m_B$ = 13 mag with $\delta > 0$ in the CfA catalogue (Huchra et al.
1989). For practical reasons only galaxies with diameters smaller than 4' were retained. To
minimize inclination corrections for both photometric and kinematical data, we only considered
objects with inclinations between 32\degr~ and 73\degr (i.e., catalogued $b/a$ in the range
0.8$< b/a <$ 0.3). The sample selected in that way is magnitude and size limited.

After that we defined a criterium for isolation. We first excluded all the galaxies that are in
the catalogues by Karachentsev (1972), Turner (1976) or Soares (1989), that is to say, well
characterized members of isolated pairs or small groups.  Then, a galaxy was considered {\bf
isolated} when the nearest neighbor found in the CfA catalogue was outside the volume defined
by a projected distance of 0.5 Mpc (we adopt here H$_0$ = 75 \kms~Mpc$^{-1}$) and a redshift
difference of 500 \kms. To eliminate the possibility of having small companions fainter than
the limit of the CfA catalogue, that could be still very efficient in producing dynamical or
morphological perturbations in the main galaxy (see quoted references and M\'arquez et al.
1996), we also excluded galaxies with optical companions in the Palomar Sky Survey Prints. The
final sample contains 22 galaxies defined as isolated (see Table 1 in Paper I). It is found
that all of them have $cz < $ 6000 \kms.

\section{The morphology of the isolated galaxies}

The RC3 catalogue (de Vaucouleurs et al. 1991) contains detailed morphological information for
many of the 22 galaxies in our sample. There is information on the subtype for all the
galaxies, with the exception of NGC~6155, classified just as spiral. There is catalogue
information on the presence or not of a bar for 16 galaxies, and we were able to add that
information for other 5 galaxies (UGC~3511, UGC~3804, NGC~3835, NGC~6155 and NGC~6395, see
notes on individual galaxies in Paper I\footnote{Published only in electronic form and
available on the server of the Editions de Physique: http://www.ed-phys.fr}), so the bar
information does exist for all the 22 galaxies but one, namely NGC~4525, classified as Scd. The
morphological information is completed with details on the presence of rings for 17 galaxies in
the sample.

Looking at that information, the following picture emerges. First, only 5 out of the 21
galaxies with type information are earlier than Sc. This could be
just an artifact due to the limited size of our sample, but we note that de Jong and van der
Kruit (1994) find also a type distribution that peaks at Sc for their sample of 86
non-interacting, non-peculiar spirals. 

Among the 21 galaxies with information on the presence of a bar, 8 are non barred (SA), 3 are
barred (SB) and 10 are weakly barred (type SAB coded as SX in the RC3). The fraction of barred
systems among the isolated galaxies amounts to 62\%. In spite of the small size of our sample,
we notice that this fraction is very similar to that found for a large population of spirals,
without consideration of their environmental status (Moles, M\'arquez \& P\'erez 1995).

Another interesting aspect is that all the non-barred (SA) galaxies in the sample are ringed
spirals. Therefore, all the 21 galaxies with bar or ring information do show the presence of
features indicative of the presence of non-axisymmetric components of the potential. Indeed,
this kind of structures is easily explained as due to perturbations of the gravitational
potential by companions (Simkim et al. 1980; Arsenault 1989; Elmegreen et al. 1990; Combes \&
Elmegreen 1993). But the selection criteria we have adopted to define the sample of isolated
spirals was designed to select objects that wouldn't have experienced gravitational interaction
in the last 10$^9$ years at least. The presence of those components in all of them should imply
longer time scales for those features, unless the possibility of spontaneous formation of such
structures along the galactic life are accepted.

The global aspect of those galaxies is however quite regular. For the 17 galaxies for which we
could gather CCD broad band images, our analysis shows that all their disks are quite symmetric
at their outskirts. We have calculated the decentering degree as the displacement of the center
of the most external, recorded isophote with respect to the luminosity center, normalized to
the last measured radius. The values we measured are always smaller than 5\%, except for NGC
6155, for which we found 10\%. The average decentering amounts to 2.4 $\pm$ 2.7 \%. Concerning
the spiral arm structure, it can be generally described as regular and symmetric in shape,
although sometimes it is more intense in one of the hemispheres. Essentially all kind of
structuring are encountered in our isolated spirals but, we cannot extract statistically significant 
results on their arm structure due to the limited number of systems we have in the sample.

\section{The photometric and kinematical properties}

In Paper I we give the whole set of parameters measured for the sample of isolated galaxies,
together with the descriptions of how they have been obtained. In table \ref{aa2_para} we show
the median values together with the dispersion and the range of variation for the different
measured parameters. Indeed, given the size of the sample it is not possible to give the values
for each morphological subtype. Moreover, as we have discussed, most of the galaxies are late
types, with 11 of them classified as Sc. We note that this aspect should act as a caution when
trying to do comparative studies.

\subsection{The photometric properties}

\subsubsection{Luminosity and Color Indices}

The distances used to evaluate the luminosity were derived from the  measured redshift
corrected for galactocentric motion (as indicated in the RC2 catalogue, de Vaucouleurs et al.,
1976), with H$_0$ = 75 \kms~Mpc$^{-1}$. The correction for the Virgocentric inflow for our
galaxies is in general small (always less than 15\%) and, as discussed in Paper I, not
sensitive to the detailed model used. We therefore decided not to correct for the inflow. The
magnitudes and color indices given in table 2 are corrected for galactic and internal
extinction as explained in Paper I (see Table 7 in Paper I).

\begin{table}[]
\caption[]{Average properties of the 22 isolated spirals in the sample.}
\label{aa2_para}
\begin{flushleft}
\begin{scriptsize}
\begin{tabular}{|l l|  l l|} 
\hline
$t$ & 5$\pm$1.4                &$-M_{B}^D$ & 19.98$\pm$0.53\\
$-M_B^0$ & 20.35$\pm$0.54      &$-M_{I}^D$ & 21.70$\pm$0.68\\
$-M_V^0$ & 20.74$\pm$0.60      &$-M_{B}^B$ & 17.80$\pm$0.80\\
$-M_I^0$ & 21.88$\pm$0.85      &$-M_{I}^B$ & 20.44$\pm$0.93 \\
$(B-V)_T$ & 0.48$\pm$0.09      &$(B-I)_D$ & 1.75$\pm$0.24 \\  
$(B-I)_T$ & 1.67$\pm$0.25      &$(B-I)_B$ & 2.52$\pm$0.59 \\  
$<\mu>$& 21.57$\pm$0.28        &$(B-V)_c$ &0.89$\pm$0.10\\ 
$R_B^e$ & 4.3$\pm$1.4          &$(B-I)_c$ & 2.21$\pm$0.29 \\
$R_V^e$ & 3.5$\pm$1.2          &$-P_{(BV)}$ &0.047$\pm$0.019\\  
$R_I^e$ & 3.2$\pm$1.0          &$-P_{(BI)}$ &0.077$\pm$.033\\
$D$ & 35$\pm$12                &$G$ & 157$\pm$103 \\       
$i$ & 53$\pm$10                &$R_{max}$ (Kpc)& 4.00$\pm$1.28\\     
$a$ & 24.4$\pm$9.6             &$V_{max}^{i}$ & 179$\pm$35\\    
$b$ & 12.7$\pm$6.7             &$R_{M}$ (Kpc)& 8.0$\pm$2.6 \\      
$\mu_{B,D}^e$ & 22.61$\pm$0.44 &$V_{M}^{i}$ & 183$\pm$35\\     
$R_{B,D}^e$ & 5.1$\pm$1.5      &$Mass(R_M)$ & 7.5$\pm$3.9\\      
$\mu_{B,B}^e$ & 20.10$\pm$2.25 &$Mass(R_{25})$ & 10.3$\pm$6.1\\    
$R_{B,B}^e$ & 0.22$\pm$0.65     &$V_{sist}$ & 2569$\pm$859\\       
$\mu_{I,D}^e$ & 20.40$\pm$0.54  &$M/L_B$ & 4.1$\pm$1.4 \\
$R_{I,D}^e$ & 3.6$\pm$1.1       &$\Delta^{(1)}$ & 7.6 $\pm$ 7.1\\       
$\mu_{I,B}^e$ & 17.96$\pm$2.35  &$log(L_{H\alpha}^{(2)})$ & 41.2$\pm$0.5\\ 
$R_{I,B}^e$ & 0.73$\pm$1.02 & ~ & ~\\   
$0.5 \times D_{25}^B$ & 10.01$\pm$2.85 & ~ & ~\\   
$0.5 \times D_{25}^I$ & 12.93$\pm$3.52    & ~ & ~\\
\hline
\end{tabular}
\noindent

\smallskip
M are absolute magnitudes. 
All the R (effective radius) together with $a$ and $b$ (major and minor axes) 
are in Kpc. 
D are distances in Mpc (H$_0$=75 \kms Mpc$^{-1}$). 
$i$ = acos ($b$/$a$), in degrees. 
$\mu^2_{i,j}$ are surface brightness parameters in 
mag/('')$^2$, and $R^e_{i,j}$ the  effective radii in Kpc, from the 
photometric decomposition, where $i$ is the filter and $j$ is the 
component ($D$=disk, $B$=bulge). $O$ and $P$ are the origin and slope of the 
linear regime of the color gradients (see text). 
G is in units of \kms Kpc$^{-1}$.
M/L$_B$ is in solar units. Velocities are in \kms.  
Mass in 10$^{10}$M\sun~ units\\
$^{(1)}$ See text for the definition of $\Delta$ = $\Delta V/\Delta R$ (\degr)\\
$^{(2)}$ Luminosity in 10$^{41}$erg/s units.\\
\end{scriptsize}
\end{flushleft}

\end{table}

The range of the B luminosity of the galaxies we have measured is $-18.58\leq M_B\leq -22.19$,
with a median value of $-20.35$ (see table 1). Therefore, there are no faint spirals in our
sample, a fact that has to be taken into account when making comparative analysis (see below).
The effective radii in Table 1 have been calculated from the growing curves in the different bands.
The mean surface brightness values were evaluated for the area enclosed by the 25 mag/('')$^2$ isophote 
(taken from the RC3), i. e., $<\mu> = -2.5 log (I/r_{25}^2)$, where $r_{25}$ is the radius of that 
isophote.

For the total color indices the ranges we find are $0.35\leq (B-V) \leq 0.85$ and $1.2 \leq
(B-I) \leq 2.4$. These are similar to what is found for other samples of spiral galaxies,
independently of their interaction status (Roberts \& Haynes 1994; de Jong 1996c).

\begin{figure*}
\centerline{
\psfig{figure=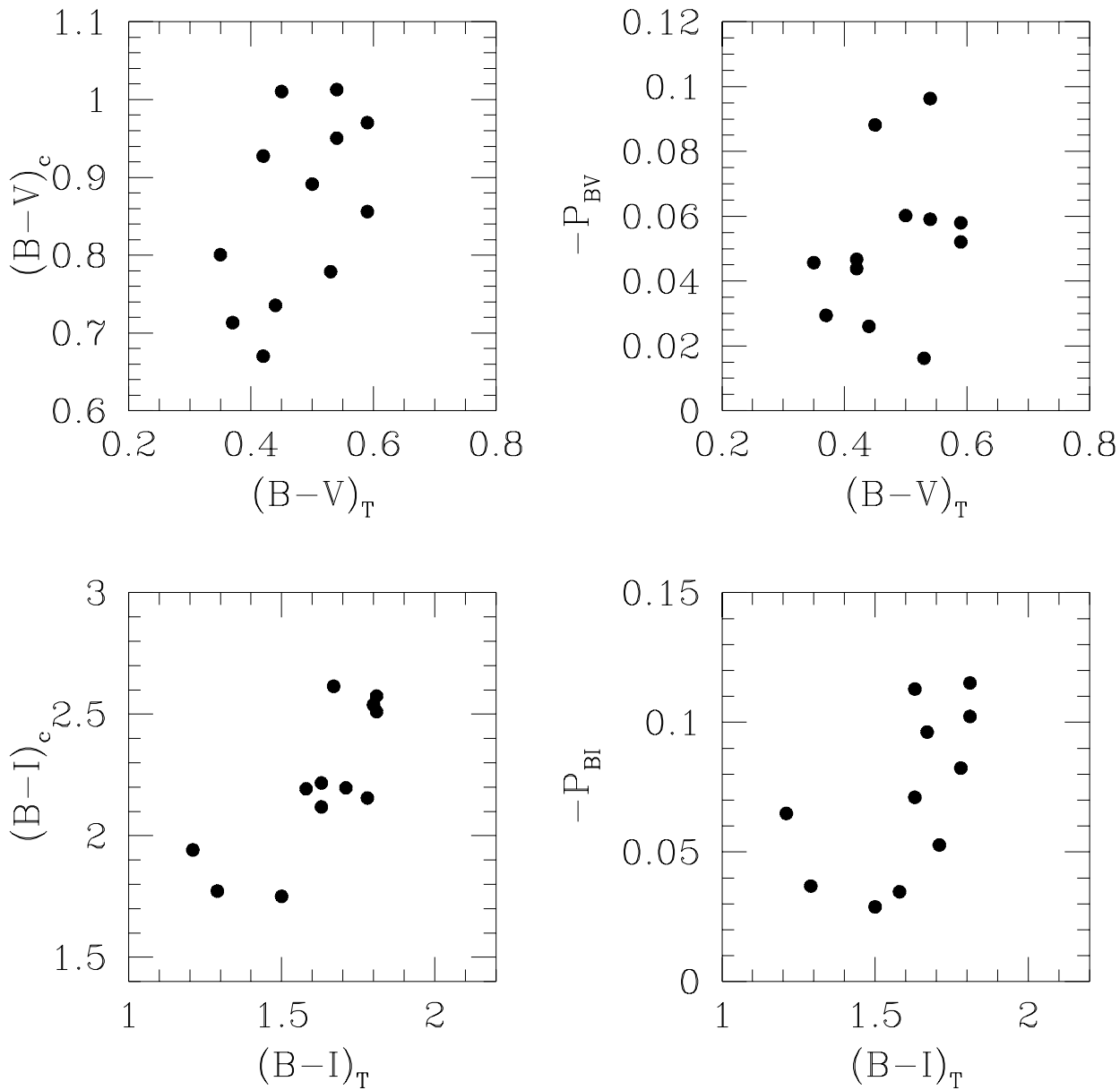,width=17truecm}}
\caption[]{The central colors, and the slope P of the color gradient (in magnitudes per kpc)
as a function of the total galactic colors.}
\protect\label{colores}
\end{figure*}

To characterize the populations of the disks we have considered the color indices and their
gradients along the disk. To quantify the color gradients we have fitted a linear function of the form 
CI(r) = CI(c) + Pr, where CI(c) is the (extrapolated) central color index under consideration, and r the 
radial distance in kpc. We find that P is negative for all the galaxies in the sample, i.e., their color 
indices become bluer towards the external parts. As shown in Figure \ref{colores}, the central color 
indices correlate with the corresponding total colors: they are redder for redder galaxies. The 
correlation is better traced by $(B-I)$, with correlation coefficient r = 0.774 and probability 
P = 0.9969. Another interesting aspect is that the central colors seem to span a wider range
that outer colors, specially in (B-I). We find $0.67\leq (B-V)_c\leq 1.01$ and $1.75\leq (B-I)_c
\leq 2.61$ for the central colors, whereas for the colors measured at the last recorded
isophote, R$_B$, we have $0.50\leq (B-V)\leq 0.78$ and $1.52\leq (B-I) \leq 2.17$ (UGC~3511 was
excluded for this calculation since the colors we measured are abnormally red for a Scd spiral,
see Paper I). 

The emerging picture from the above considerations is that redder galaxies have redder central colors. The fact 
that the range of color indices at the outer regions is smaller than at the center would mean that the 
disks of different galaxies tend to be more alike that their bulges.

\subsubsection{The properties of the bulges and disks}

It is a well known fact that the output of the photometric decomposition of the light
distribution in the image of a galaxy depends on the method used and on the form of the
profiles adopted for the components (Knapen and van der Kruit, 1991). To make explicit our
choices we notice that we have used 1-D light profiles, with an exponential law for the disk
(Freeman 1970) and the r$^{1/4}$ law for the bulge (de Vaucouleurs 1948), respectively:

$$ log(I_D^i) = log(I_{Do}^i) - 0.7290 (r/r_D^i -1)$$
\noindent
and

$$ log(I_B^i) = log(I_{Bo}^i) - 3.3307 ((r/r_B^i)^{1/4} - 1).$$

\noindent
where i stands for the photometric band under consideration.

The isophotal profiles have been derived by plotting the isophotal levels {\it versus} their
equivalent radii, calculated from the area inside each observed isophote. Disk and bulge
parameters have been obtained from the surface brightness profiles, 
following Boroson (1981)
and using the marking the disk method.

\begin{figure*}
\centerline{
\psfig{figure=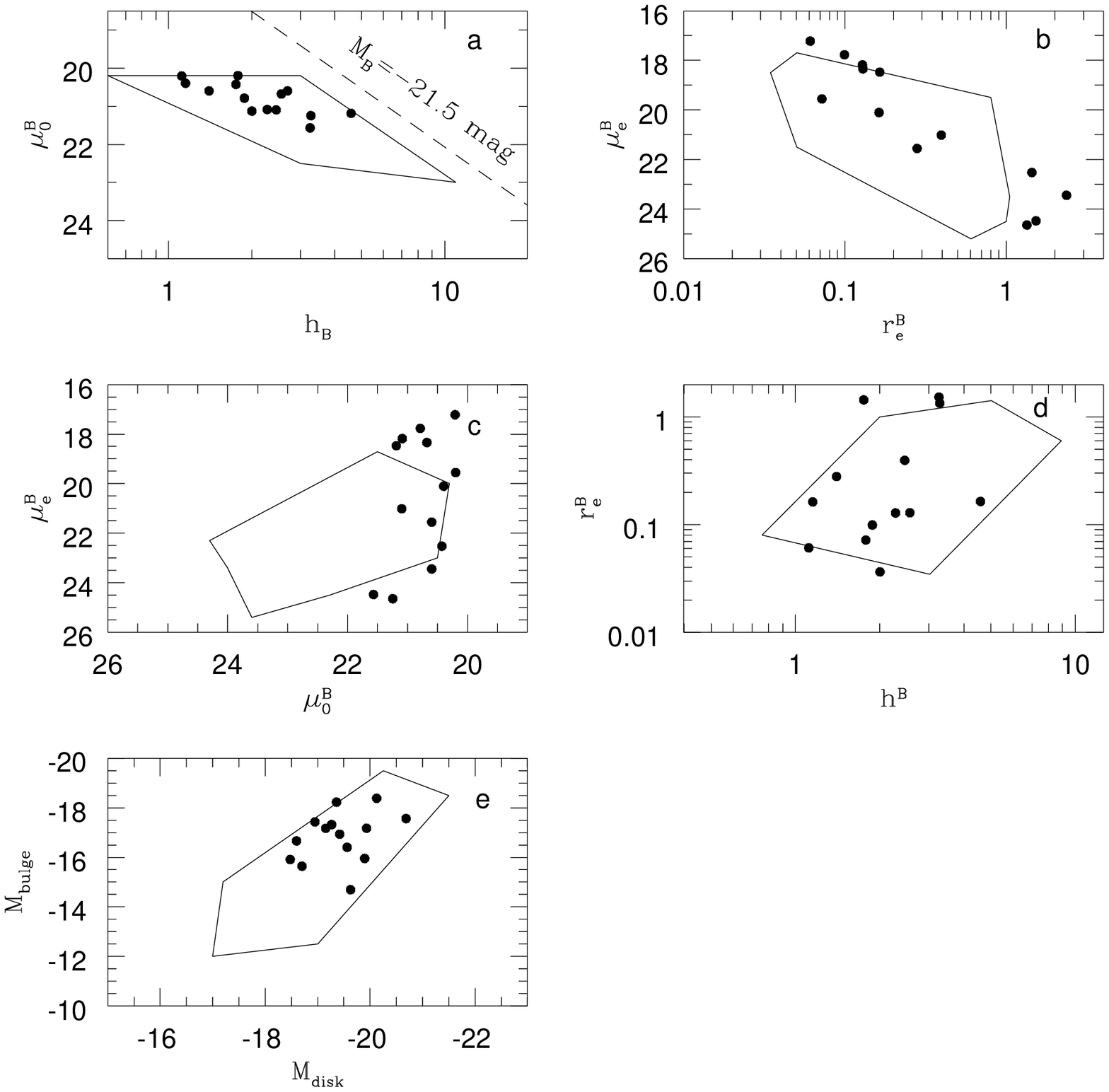,width=14cm}}
\caption[]{The bulge and disk properties of the isolated galaxies in the B band. $\mu_0^B$ {\it versus} $h_B$ is given in (a).
Panel (b) is the Kormendy relation for the bulges of isolated spirals. The bulge {\it versus}
disk surface-brightness, scale length and luminosity are shown in panel (c), (d) and (e),
respectively. In the figures we
have also plotted schematically the regions covered by the data obtained by de Jong (1996b) for
non-interacting, non-perturbed spirals,
represented by closed boxes.}
\protect\label{dJIII}
\end{figure*}

\begin{figure*}
\centerline{
\psfig{figure=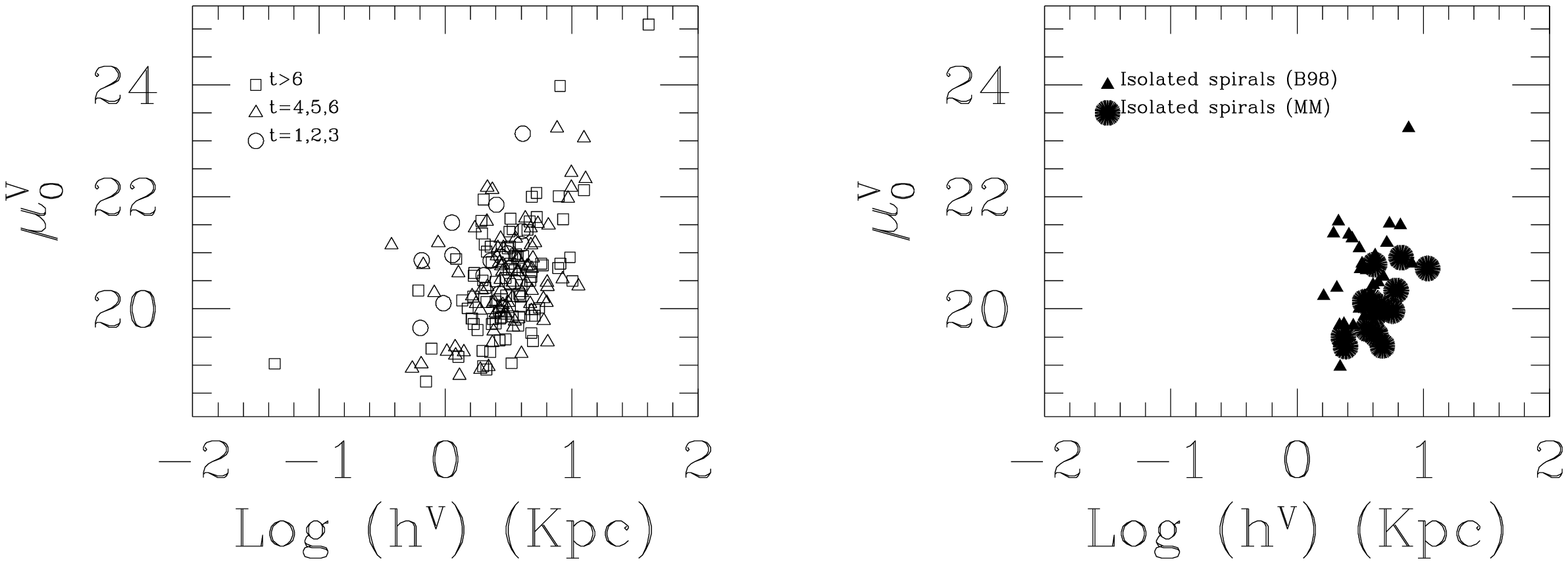,width=17.5cm,angle=-90}}
\vspace{-0.6truecm}
\centerline{\psfig{figure=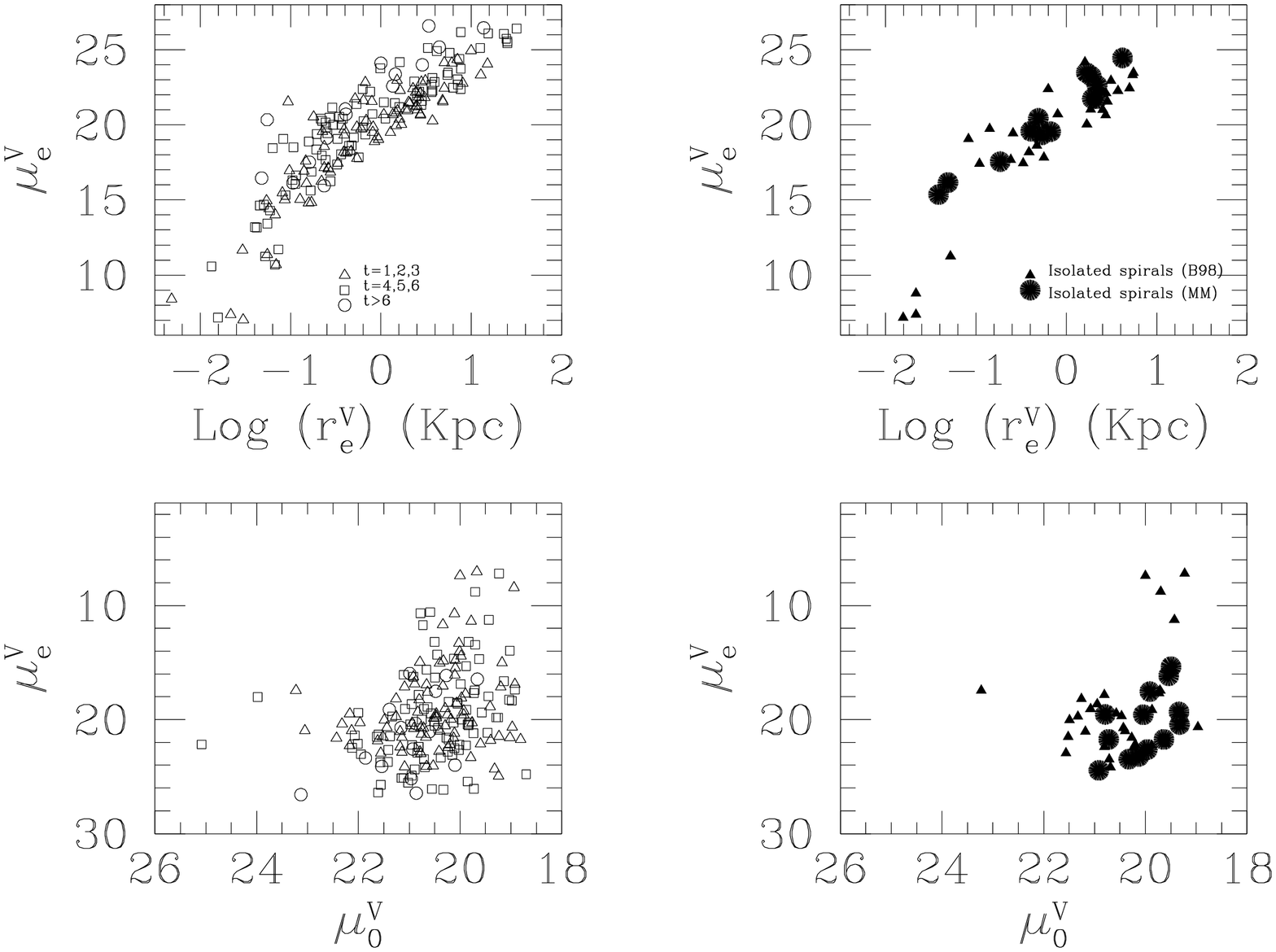,width=17.5cm,angle=-90}}
\vspace{-0.9truecm}
\centerline{\psfig{figure=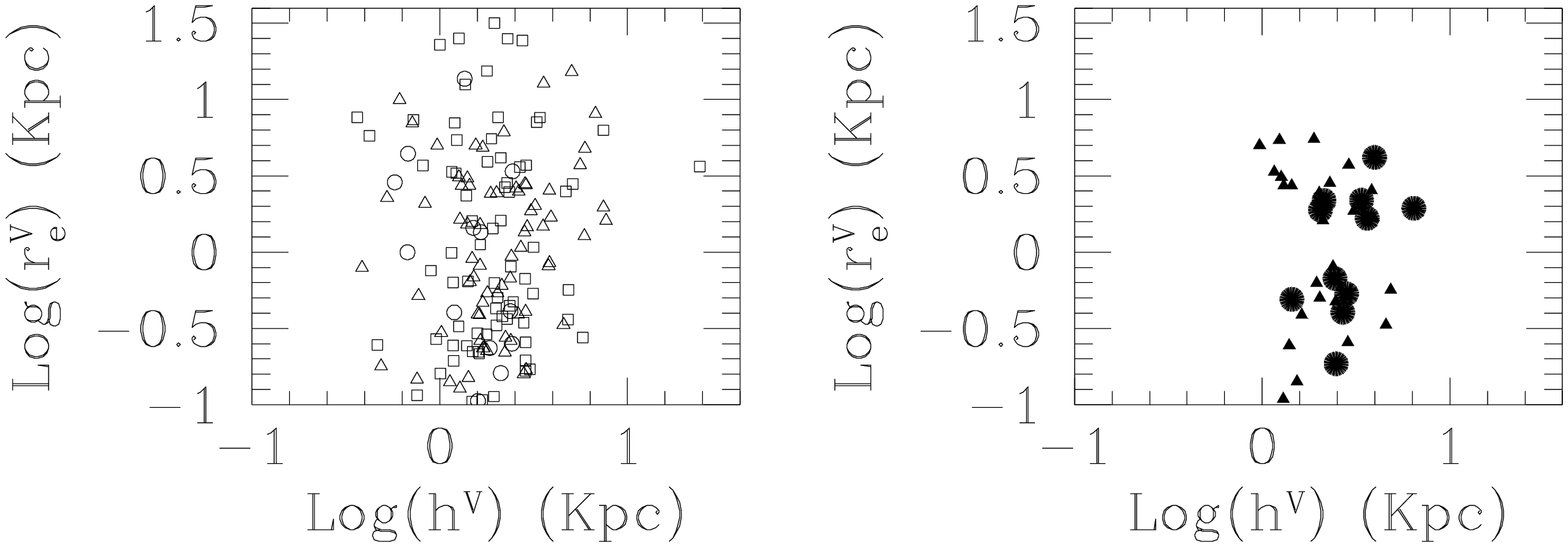,width=17.5cm,angle=-90}}
\vspace{-0.5truecm}
\caption[]{Relations for bulge and disk parameters.
In the left panels we plot data from all spirals with type I profiles from Baggett et al.
(1998) (196 galaxies). Sa to Sb are plotted as open triangles, Sbc to Scd as open squares and
later than Scd as open circles. In the right panels we only plot the galaxies selected from
Baggett et al. as isolated (filled triangles), together with our sample spirals (dark
circles).}
\protect\label{baggett}
\end{figure*}

The main results are given in table 1, and presented in the different panels of Figures
\ref{dJIII} and \ref{baggett}. (NGC~718 data appears as a discrepant point in all the relations
involving its B-magnitude. We suspect that the abnormally red colors we have measured are not
correct and the galaxy should be observed again before being included in the discussion. This
is the reason to omit it in the following.) No trend is found between the disk and bulge
parameters and the morphological type. In particular, for the Scs in our sample it is clear
that their disk and bulge properties span a big range. Indeed, the size of our sample is too
small to draw conclusions. But the trend we find for the isolated galaxies is much alike to
that shown by larger samples of non-interacting galaxies. We have to insist, before
starting comparisons between different sets of data, on the differences that can be induced by
the use of different methodologies. Thus, it has been argued that exponential rather
than r$^{1/4}$ fits would be more appropriate for the bulges of late spirals (Andreadakis \&
Sanders 1994; de Jong 1996a; Courteau, de Jong
\& Broeils 1996, Seigar \& James 1998). The resulting bulges are then fainter than when a
r$^{1/4}$ law is fitted. Moreover, the use of 1-D bulge profiles (our case) produce bulges with
fainter central surface brightness and larger effective radii than 2-D fits.

With all this in mind, we can compare our results with the B-band data presented by de Jong
(1996b) for a sample of non-perturbed, non-peculiar spiral galaxies, a sample that, as we
already argued, can be taken as not too disimilar to ours except in the luminosity range. In the same 
figure \ref{dJIII} we
also present de Jong's data. It is clear that there is a large overlap between both sets of
results.

The slight differences that can be appreciated after a more detailed look, can be explained in
terms of the differences in methodology and the already quoted bias towards luminous galaxies in our 
sample.
Thus, the mean value of the central disk surface brightness that we find is $\mu_B^0$ = 20.9
$\pm$ 0.6, rather on the bright end of the values determined by Bosma \& Freeman (1993) and by
Giovanelli et al. (1994), and higher than the value given by de Jong. 
The same is found for the derived disk luminosities.
This is certainly
due to the absence of galaxies fainter than M$_B = -$18 in our sample since, as pointed out by
de Jong (1996b) and Courteau (1996), the average value of the central surface brightness of the
disks (Freeman 1970) depends on the luminosity range considered. Similar considerations apply
to the bulge parameters we have derived.

We find that the scale length values for the disks depends on the photometric band
(see table \ref{aa2_para}), in the sense that it becomes smaller for redder wavelengths. Evans
(1994) has argued that this would be due to the effect of dust layers in non-transparent disks,
but de Jong (1996c), who found a similar result to ours, was able to model this behaviour by
combining the presence of disk gradients in both, stellar age and metallicity.

On the other hand, the relations between the surface brightness and scale parameters for both
components appear to be better correlated for our galaxies than for de Jong's data (see panels
a - disk - and b - bulge - of figure \ref{dJIII}). The Kormendy relation for the bulge is
significantly tighter for our data (in fact, de Jong reported no correlation between $\mu_e$ and $r_e$ 
for his data). A similar relation appears for the disk, with a scatter much lower than previously found.

The relation between the corresponding parameters of the bulge and disk components are
presented in the panels c, d and e in figure \ref{dJIII}. It can also be seen that the range of values 
spanned by the disks is significantly smaller than that of the bulges. This would indicate that the spiral galaxies differ between them mainly by the bulge properties, the disks being much more similar.

We have also compared our results with the data presented by Baggett et al. (1998) in the V band
(figure \ref{baggett}). The comparison indicate the same trends already noticed: The relations
between parameters become tighter when only isolated or similar galaxies are considered. And,
as before, the disk parameters span a significantly smaller range than the bulge parameters.

\begin{figure*}
\centerline{
\psfig{figure=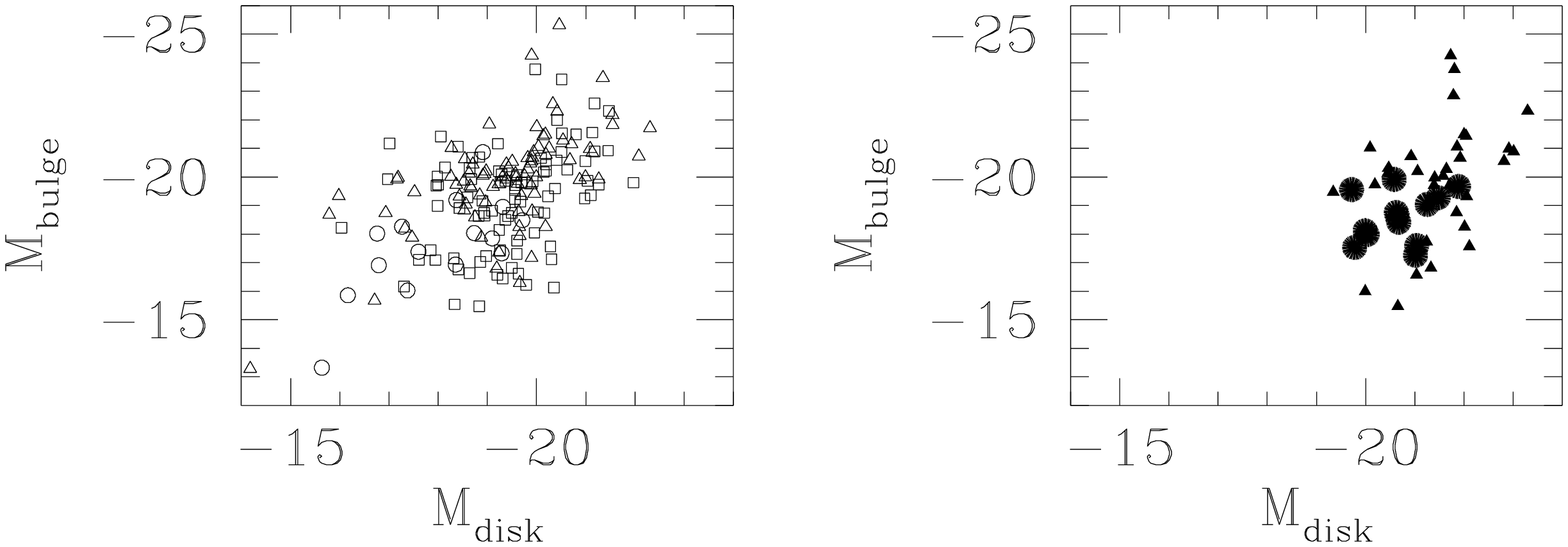,width=17.5cm,angle=-90}}
{\bf Fig. 3.} (Cont.)
\end{figure*}

Thus, the trends show by our data, in spite of the small sample we have, are supported as
physically meaningful when larger samples of galaxies in acceptably similar conditions are
considered. The data indicates that for isolated or non-interacting galaxies there is a tight
relation between the surface brightness and size not only for the bulge (Kormendy relation) but
also for the disk, and that these relations are more scattered when interacting galaxies are
added. We will discuss in Paper II whether this is related to the lack of faint galaxies in our sample 
or to the interaction status. On the other hand, the disks of different spirals, no matter their 
morphological types, are much more alike than their bulges. 

\subsection{The kinematical properties. The masses and M/L ratios}

The parameters describing the rotation curves are given in Table 10 of Paper I, whereas the
median values are given in Table \ref{aa2_para} here. We recall that the inner gradient, G (in
kms$^{-1}$kpc$^{-1}$) is defined as G = ($v_G^{ob}$/sin(i))/$r_G$, where $r_G$ (in kpc) is the
radius of the inner region of solid-body rotation, $v_G^{ob}$ (in \kms) is the observed
velocity amplitude at $r_G$, and $i$ is the disk inclination, as derived from the optical
images. Other parameters in the table are $R_{max}$ (in kpc), the radius at the point of
maximum rotation velocity, $V_{max}^i$ = $V_{max}^{ob}$/sin(i);
R$_M$, the radius at the last measured point in the rotation curve, with $V_M$ its
corresponding velocity. The mass has been evaluated at $0.5 \times D_{25}$, for a simple homogeneous and
spherically symmetrical distribution, M$_{25}$ = 2.3265 $\times$ 10$^5$ $0.5 \times D_{25}~V_{25}^2$M$\sun$
(Burstein and Rubin 1985).

Our results for V$_M$ and M$_{25}$ are in agreement with those found for Sb, Sbc and Sc
galaxies by Rubin et al. (1982). With the mass calculated as described,  we have evaluated the
M/L ratio for different bands. The median total mass-to-luminosity ratio is $M/L_B$ = 4.1 $\pm$
1.3, spanning a range from 1.7 to 7.6. For the other bands we find similar ranges and central
values, 5.5 $\pm$ 2.0 in V, and 4.5 $\pm$ 2.7 in I.

We choose the parameter $\Delta$ = $tg^{-1}$[(V$_M$$-$V$_{max}$)/(max(V$_M$$,$V$_{max}$))
 $\times$ (R$_M$$-$R$_{max}$)/($0.5 \times D_{25}$)] to describe the overall shape of the rotation curve
farther than $R_{max}$. In other words, the velocities are normalised to the maximum amplitude
and the radii to $0.5 \times D_{25}$. The parameter $\Delta$ can take values around zero (for flat
rotation curves), positive (for rising rotation curves), or negatives (for declining rotation
curves). The median value obtained for isolated galaxies, $\Delta\approx$ 8 $\pm 7 \degr$, is
compatible with flat rotation curves.

\subsection{The star formation properties}

A number of emission line regions, including all the nuclei we have observed, are present in
our long slit spectra. Due to the wavelength coverage only the [OI]$\lambda$6300, H$\alpha$, [NII]$\lambda\lambda$6548,6583  and
[SII]$\lambda\lambda$6713,6731 lines could be detected. Our analysis will mainly concentrate on the properties of the
strongest H$\alpha$ and [NII]$\lambda$6583 lines. 

The general aspect of all the spectra of the detected emission line regions, including the
nuclei, is that of normal HII regions photoionized by stars. The [SII]$\lambda\lambda$6713,6731
line ratio is always $\geq$ 1, indicating low electronic densities, as expected for such
regions.

The median
value of EW(H$\alpha$) is 11\AA~ and 18\AA~ for the nuclei and the external 
regions, respectively.
The
[NII]$\lambda$6583/H$\alpha$ line ratio ranges from 0.12 to 0.8 for all the nuclei with
EW(H$\alpha$)$\geq$2\AA, with a median value of 0.49. Indeed, for the 3 nuclei with
EW(H$\alpha$)$<$2\AA, the line ratio is very high, but this is due to the fact that the measure
of the H$\alpha$ line intensity is severely affected by the underlying absorption.

For 104 of the 105 non-nuclear HII regions detected in our major axis long slit spectra the [NII]$\lambda$6583/H$\alpha$ line ratio could be measured. The median value is 0.38, with 75$\%$
of the cases between 0.30 and 0.50. There are 4 extreme cases, with very high values, that are due
to the presence of relatively strong absorption under H$\alpha$.

Thus the central values for nuclear and external HII regions are rather similar, with a larger range 
for the external regions. These results will be considered in more detail in Paper III, 
where they will be compared with data for HII regions in interacting galaxies.

Concerning the results from H$\alpha$ CCD photometry, we measured 7 isolated galaxies, whose
total H$\alpha$ fluxes are given in Table 7 of Paper I (we have excluded NGC~718 from this
analysis, since our spectroscopic data show that the contamination from [NII] emission lines is
very important, see Paper I). The median value for the total H$\alpha$ 
luminosity 
is log(H$\alpha$) = 7.89 $\pm$ 0.55 (in solar luminosities), 
that is well within the range found for spirals (Young et
al. 1996, and references therein). The H$\alpha$ luminosities are well correlated with both the
optical area of the galaxies and their FIR luminosities. From the first relation we derive
H$\alpha\propto S$. This would mean than the average star formation per unit 
area, as judged from the H$\alpha$ luminosity is about the same in all isolated spiral galaxies.

\begin{figure}
\centerline{
\psfig{figure=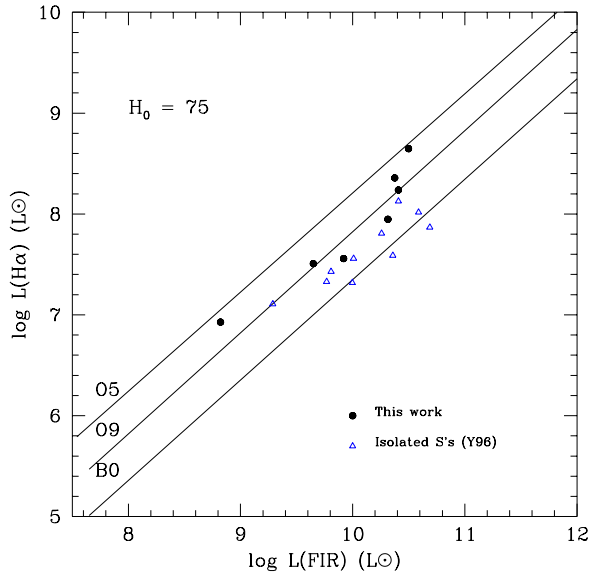,width=8cm}}
\caption[]{Relationship between integrated H$\alpha$ and FIR luminosities. Solid dots are for our
isolated spiral galaxies, triangles are for the sample of isolated galaxies by Young et al.
(1996). The solid lines represent the expected relations for O5, O9 and B 0 stars, as presented
by Devereux \& Young (1990)}
\protect\label{young8}
\end{figure}

The FIR luminosity was calculated as in Young et al., $L(FIR) = 3.75 \times 10^{5}
D^2C[2.58 S(60) + S(100)]$, where $D$ is the distance to the galaxy in Mpc, $C$ is a correction
factor for the flux longwards of 120 $\mu$m and shortwards of 40 $\mu$m (they depend on
$S(60)$/$S(100)$ and are taken from Table B.1 of the Catalogued Galaxies in the IRAS Survey,
Londsdale et al. 1985) and $S(60)$ and $S(100)$ are the IRAS 60 and 100 $\mu$m fluxes in Jy. In
figure \ref{young8} we plot the comparison of H$\alpha$ and FIR luminosities for our sample
galaxies, together with the points corresponding to the 10 galaxies catalogued as isolated by
Young et al. Both samples of isolated spirals agree within the errors. They are all in the
region occupied by normal spirals, compatible with ionization produced by O5 to B0 stars. We
anticipate here that this is not the case for all the spiral in isolated pairs (Paper III).

Finally, we have also analysed the star formation history of those galaxies for which we have a
complete set of data, i.e., the B-luminosity, the H$\alpha$ integrated flux, and the total
mass. Following Gallagher et al. (1984), we quantify the present star formation rate in terms
of the H$\alpha$ luminosity, the
star formation rate during the past 10$^9$ years as a function of the B luminosity, and the initial star formation rate as a function of
the total mass. Unfortunately we have all the relevant data for only 5 galaxies in our
sample. Still, the result is compatible with a constant star formation rate along their
lifetimes, with a smooth SFR, for the five galaxies. We will report in Paper III that this is
not the case for some of the spiral galaxies in isolated pairs.

\section{The relations between the measured parameters}

It is well known that the observational parameters describing spiral galaxies are correlated,
in such a way that two quantities can describe most of the variance in the parameter space. In
his pioneering analysis, Brosche (1973) used the method of the Principal Component Analysis
(PCA) to study the data (morphological type, optical size, color, absolute luminosity, maximum
rotation velocity and HI mass) of 31 spiral galaxies. He found that the parameter space has two
significant dimensions. Bujarrabal et al. (1981) also used the PCA to study a sample of
$\approx$ 100 objects with optical and radio data and also conclude that 2 parameters could
suffice to describe the data: {\it size} (given equivalently by the optical size, the HI mass
or the luminosity) and {\it aspect} (given by the morphological type or the color). Whitmore
(1984) obtained the same conclusion for a sample of 60 spirals with about 30 observed
parameters, and assigned the two dimensions to {\it scale} (optical size, luminosity) and {\it
form} (color, bulge to total light ratio). More recently, the PCA of I-images of some 1600
spirals led Han (1995) to find also two principal dimensions. Magri (1995), with a sample of
492 spirals with compiled and new data (Magri 1994), and a refined version of the PCA to allow
for the inclusion of non-detections, also concluded in the same vein.

Even if the PCA has some limitations related with the bias in the results that could be
introduced if the number of parameters involved in the analysis is not big enough (Magri 1995),
or the fact that it assumes that the correlations among the parameters are strictly linear, it
appears as a well suited method to find the minimum number of variables describing the data set
we have constructed for isolated spiral galaxies. In our case, given that all possible effects
of the interaction are in principle excluded, what could be expected from that kind of analysis
is to obtain the bare, intrinsic structural correlations between the parameters. We have
already discussed how the inclusion of interacting galaxies could increase the scatter in the
relations between parameters. In the same sense, Folkes et al. (1996) have shown that the
inclusion of perturbed or peculiar galaxies could produce misleading results, in the sense of
smoothing the morphology-spectrum relationship found for normal galaxies.

\begin{figure}
\centerline{
\psfig{figure=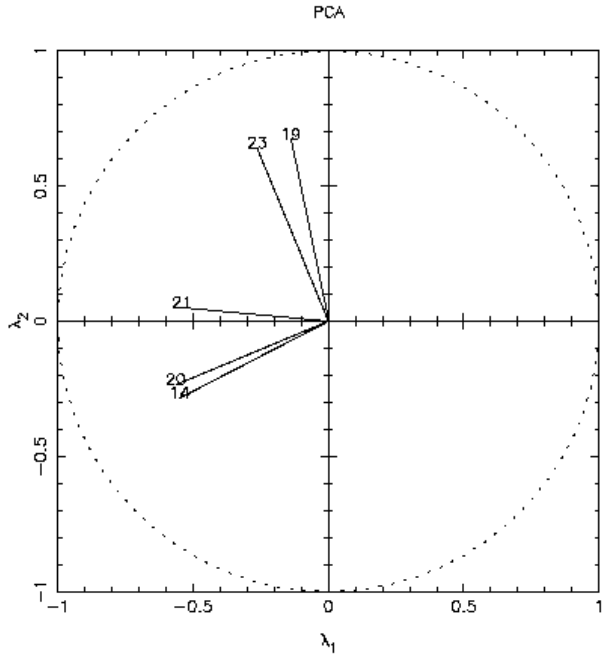,width=8cm}}
\caption[]{Results of the PCA. The vectors are 14=G, 19=L$_B$, 20=
B/D=, 21=(B$-$V), 23=Mass.}
\protect\label{pca}
\end{figure}

The main result that emerge from the PCA of our data is that the parameter space has
essentially 2 dimensions, in agreement with all the previous results. The new aspects of our
analysis are that till 95 $\%$ of the variance of the sample is explained, and the
identification of a better suited form parameter. Indeed, the first {\sl eigenvector},
corresponding to the {\it scale} parameter, is given by either the total luminosity, L$_B$, the
total mass, M, or the size, as usual. For the second, the {\it form} parameter, we find that
the best choice is the inner gradient, G, or, equivalently, the bulge to disk ratio, B/D. In
figure \ref{pca} we show the projection of the vector constructed with the luminosity, the
bulge to disk ratio, the total mass, the G parameter and a color index onto the plane described by the 
two eigenvectors. We note that we find no correlation between the color and the luminosity, 
whereas color is correlated with the bulge-to-disk ratio, in the expected sense of redder galaxies having 
a greater relative contribution from the bulge component.

It is the first time that the G-parameter is revealed as equivalent to the form {\sl
eigenvector}. Some other were proposed, like some color index, or the morphological type, but
what our data on isolated galaxies show is that the scatter is minimum when G, or the B/D
ratio, is used. The general trends between the morphology, the colors and the B/D ratios are
well known, and at the base of the Hubble classification scheme. In that sense, Whitmore (1984)
pointed out that the inner gradient of the rotation curves could be a good discriminant of the
central density, i.e., the importance of the bulge component. On the other hand,
Baiesi-Pillastrini (1987) already noticed that the G-parameter is not directly correlated with
the morphological type, so the relation between the B/D ratio and the morphology would present
an important scatter, as it is. What we find here is that the G-parameter is a much more useful
property to classify a spiral galaxy, in the sense that it is essentially coincident with one
of the 2 {\sl eigenvectors} of the parameter space, i.e., normal to the size parameter (see
figure \ref{pca}).

Indeed, a number of strong linear correlations between parameters does exist. Regarding the
photometric data, we have already pointed out that the optical size and the luminosity in each
band are well correlated. The disk and bulge parameters, that have been discussed in the
previous section, also present good correlation, the tightest being between the surface
brightness and the size, both for the bulge (the Kormendy relation), and for the disk. For the
bulge parameters, we find $I_e \approx r_e^{0.64}$, i.e., somewhat smaller slope but compatible
with that of the Kormendy relation for other spiral bulges (Andredakis et al. 1995; Hunt et al.
1998) and for ellipticals (Bender et al. 1992).

\begin{figure}
\centerline{
\psfig{figure=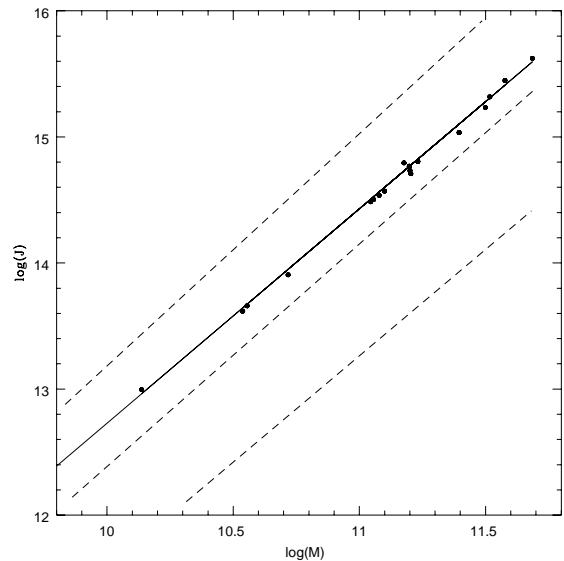,width=8cm}}
\caption[]{Correlation between the angular momentum, $J$ (see text) and the total mass, M. The
dashed lines represent the average and range found in Campos-Aguilar et al. (1993). The solid
line corresponds to the fit to our data, log($J$) = -4.27 + 1.70 $\times$ log(M).}
\protect\label{figcampos}
\end{figure}

\begin{figure}
\centerline{
\psfig{figure=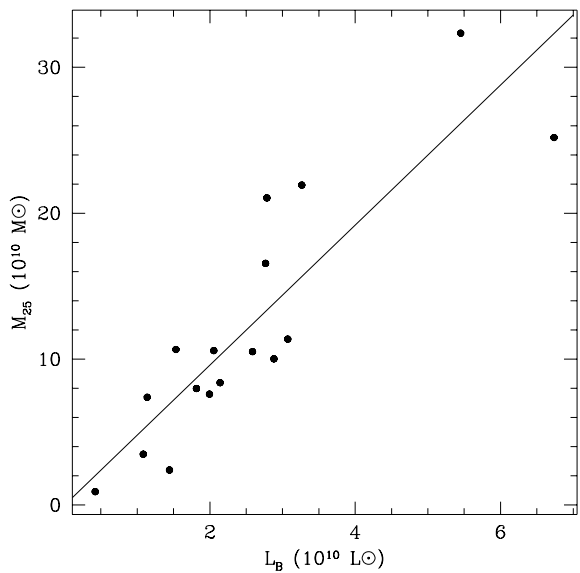,width=8cm}}
\caption[]{Correlation between the mass inside $0.5 \times D_{25}$, M$_{25}$ (see text) and the total
luminosity in B, L$_B$.}
\protect\label{ml_aa2}
\end{figure}

Among the relations involving kinematical parameters, we find a correlation between V$_M$ and
the size, in agreement with the linear relationship found by Zasov \& Osipova (1987). V$_M$
(and V$_{max}$) are also related with the B luminosity (Rubin et al. 1982; Persic \& Salucci
1991). We also find a good correlation between the angular momentum, $J$ = M $\times$ V
$\times$ R, and the total mass, log($J$) = -4.27 ($\pm$ 0.24) + 1.70 ($\pm$ 0.02) $\times$
log(M), in good agreement with Campos-Aguilar et al. (1993). The
scatter is smaller in our case (see Fig. \ref{figcampos}). V$_{max}$ and R$_{max}$ are correlated with the total magnitude
with the same overall tendencies found by Courteau \& Rix (1997) (they use R magnitudes from
Courteau 1996) but, at variance with them, we do not find any clear trend between total colors
and total luminosity, what could be due to the limited luminosity and subtype range covered by our sample. Total absolute luminosity and total mass are well correlated, as shown in
Fig. \ref{ml_aa2}. We obtain M = 4.80 ($\pm$ 0.15) $\times$ L$_B$ with a correlation
coefficient $r$ = 0.9. This implies a mass to luminosity ratio which is within the range of
(M/L) values (from 3.9 to 6.6 in solar units, with H$_0$ = 75 \kms~ Mpc$^{-1}$ ) measured for
Sc galaxies by Rubin et al. (1985). It is also compatible with (M/L$_R$) ratios found for field
galaxies by Forbes (1992) (from 3 to 6 in solar units, with H$_0$ = 75 \kms~ Mpc$^{-1}$). M/L
ratio tends to be higher for more massive (or more luminous) galaxies, as found by Broeils \&
Courteau (1996).

\begin{figure}
\centerline{
\psfig{figure=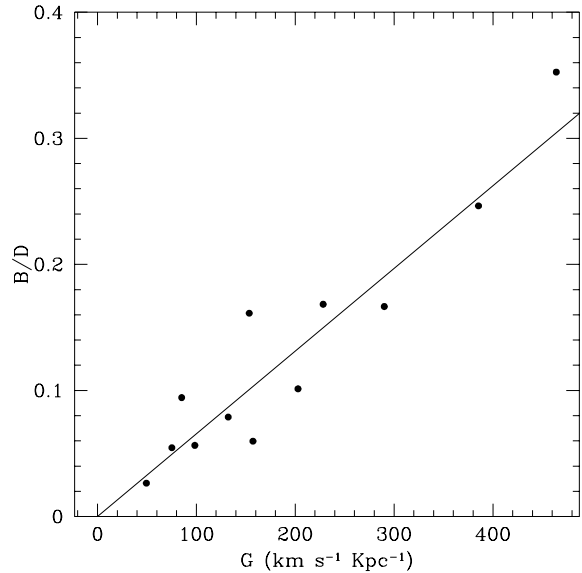,width=8cm}}
\caption[]{Correlation between the inner gradient of the rotation curve, G (\kms~Kpc$^{-1}$, and
the bulge to disk luminosity ratio, B/D.}
\protect\label{gbd}
\end{figure}

The new result is the tight correlation between the inner gradient, G, and the bulge to disk
ratio, B/D (figure \ref{gbd}) we have found. The slope is (6.80 $\pm$ 0.16)$\times$10$^{-4}$
and the scatter amounts only to 0.02 in the determination of B/D. As already mentioned, either
of those two parameters builds up the second {\sl eigenvector} of the parameter space of the
isolated spiral galaxies. The existing correlation reflects what could be intuitively expected,
that is, that galaxies with bigger bulges have higher inner gradients. Baiesi-Pillastrini
(1988) found a somewhat similar correlation but much more scattered. This could be due to the
inclusion of non-isolated spirals in his sample since, as we show in the next Paper III of the
series, the correlation is affected by the interaction, as it is much weaker for spirals in
pairs.

The correlation between G and the B/D ratio can in principle be used for distance
determination. We notice however that, as just mentioned, the scatter of the relation increases
when non isolated galaxies are considered, so only strictly isolated galaxies should be used.

\section{Conclusions}

We have characterized a sample of spirals selected as being isolated, in order to use it for
comparisons with the analogous properties of spirals in different environmental situations. We
caution that we have selected bright galaxies (brighter than M$_B = -$18.5) and that the sample is 
too small to attempt the analysis by morphological types. We further
notice that the sample is dominated by Sc type galaxies. Whether this is due to the small size
of the sample or it reflects the nature of isolated spiral galaxies should be further analyzed
with larger samples. We point out that similar result are obtained when existing larger samples
of non-perturbed galaxies are examined, so the possibility that there is a preference for later
types in poor environments should be explored.

We have used total parameters, as luminosity, color indices and mass, together with others
describing their photometrical or kinematical behaviour in more detail, as the bulge/disk
ratio, the color gradients or the shape of the rotation curve. The isolated spirals we have
analyzed can be described as following:

\begin{itemize}

\item{} The overall morphological aspect is quite symmetric and regular, as expected for galaxies
supposed to be free of external influences in the last 10$^{9}$ years at least. On the other
hand, the fraction of barred galaxies is similar to what is found for spirals irrespective of
their environmental situation. Moreover, all of them show the presence of features indicative
of non-axisymmetric components of the gravitational potential. Therefore, one is lead to
conclude that either the life time of these features is significantly over 10$^{9}$ years or
that they can be spontaneously formed along the life of a galaxy.

\item{} The total and central color indices are well correlated: redder galaxies have also
redder central regions. The outer parts of the disks are bluer and more similar than the central parts.

\item{} All of the galaxies in our sample without exception show Type I photometrical profiles.
As we will show in Paper III, this is not the case for non-isolated spirals, that can present
Type II profiles. The range of the disk scale lengths and effective surface brightness seems to
be narrower for isolated spirals than for samples including galaxies in other situations. The
disks of different isolated galaxies are bluer than their central regions, and much more alike
that the bulges.

\item{} Their current star formation rates as given by total H$\alpha$ (or FIR) luminosities are
within the range found for spirals classified as normal, i.e., without apparent peculiarities
in their morphology or nucleus. The line ratios measured in the observed emission line regions
indicate that they are photoionized by stars. For the cases for which the history of the star
formation could be retraced, it is found compatible with a smooth, constant SFR along their
lifetimes.

\item{} The overall shape of their rotation curves from R$_{max}$ to R$_{M}$ can be described as
flat. The median value of the slope in that region is found to be 8$\pm 7\degr$.

\item{} We have applied the PCA to the data set we have constructed. In agreement with previous
studies on larger samples including non isolated galaxies, we find that the 
isolated spirals in our sample 
constitute a family that can be described by two main dimensions given by size (either the
luminosity, the optical size or the total mass) and form (either the inner gradient of the
rotation curve, the G-parameter, or the bulge-to-disk luminosity ratio, B/D). The remaining
variance could just be due to the errors in the parameters. We conclude that G or B/D are more
robust and objective parameters than the morphological subtype or the color indices for the
classification of spiral galaxies.

\item{} Previous bi-variate relations are confirmed, as those between the mass and the luminosity,
the size or the angular momentum.

\item{} We have found for the first time a very tight correlation between G and B/D. B/D is a
distance independent parameter, whereas G is not. We point out that the correlation G-B/D,
constitutes a new distance indicator. We stress however that it would only be well suited for
isolated systems since, as illustrated in Paper III, the scatter of the relation significantly
increases when interacting galaxies are included.

\item{} The isolated spiral galaxies define very tight structural relations. The bulges define the
Kormendy relation with a small scatter. A similar relation is found for the disks, with a lower
scatter than in previous studies. The scatter in those relations, or in the G-B/D relation, increases 
significantly when interacting galaxies are included. Whether this is related to the inclusion of fainter 
galaxies or to the interaction status will be discussed in Paper III.

\end{itemize}

\begin{acknowledgements}

We acknowledge financial support by the Spanish DGICYT, under the program PB93-0139. One of us,
IM, acknowledges a grant from the Spanish Ministerio de Educaci\'{o}n y Ciencia, EX94~08826734,
and the hospitality of the Institut d'Astrophysique de Paris, where the writing of this paper
was started.  The Isaac Newton and Jacobus Kaptein Telescopes are operated on he island of La
Palma by the Royal Greenwich Observatory in the Spanish Observatorio del Roque de los Muchachos
of the Instituto de Astrof\'{\i}sica de Canarias.  This research has made use of the NASA/IPAC
EXTRAGALACTIC DATABASE (NED) which is operated by the Jet Propulsion Laboratory, Caltech, under
contract with the National Aeronautics and Space Administration.

\end{acknowledgements}

\noindent\null\par

\end{document}